\documentclass[twocolumn,showpacs,prl]{revtex4}
\usepackage{epsfig,amssymb}
\usepackage{dcolumn}
\usepackage{bm}

\begin{document}

\title{High Mass Particles Near Threshold}

\author{S. Reucroft$^1$, Y.N. Srivastava$^{1,2}$, \\
J. Swain$^1$ and A. Widom$^1$}

\affiliation{1. Physics Department, Northeastern University \\ 110
Forsyth Street, \\ Boston, MA 02115, USA\\ 2. Physics Department
and INFN\\ University of Perugia,\\ Perugia, Italy\\ E-mail:
stephen.reucroft@cern.ch}

\begin{abstract}
A consequence of the Higgs mechanism is that high mass particles,
such as the $Z^0$-boson, the $W^{\pm}$-boson and the $t$-quark,
are predicted to have masses that depend on the process by which
they are produced. Thus, for example, particles produced singly
are predicted to have higher masses than those produced in a pair
near threshold. Quantitative details of this prediction are
presented and discussed within the context of the current
experimental situation.
\end{abstract}

\maketitle

In the Standard Model of particle physics,
%\cite{PeskinandSchroder}
the masses of the fundamental fermions
and bosons are generated by the Higgs mechanism.  Via this
mechanism, the particle masses result from their interaction with
an external scalar field known as the Higgs field.

In a recent paper \cite{Theory} it was pointed out that high mass
particles can be used as detectors of the Higgs field.  This
possibility arises because high mass particles constitute
significant sources of the Higgs field and the resulting
source-modified field can be detected by other high mass particles
that suffer induced mass shifts.

In this paper explicit quantitative details are presented on the
size of the effect in the cases of the $Z$ mass in the $ZZ$
threshold region, the $W$ mass in the $WW$ threshold region and
the $t$ mass in the $t\overline{t}$ threshold region \footnote{In
an obvious notation, $W$ is used to represent $W^{\pm}$, $Z$ to
represent $Z^0$ and $t$ to represent the $t$-quark.}.

An overview is also given of the current status of $W$, $Z$ and
$t$ mass determinations in different production environments.

The relevant formula from \cite{Theory} is equation (19), giving:

\begin{eqnarray}
\Delta M_X &\approx& -{\Gamma_X} \left(\frac{M_X^2}{2\pi
v^2}\right) \nonumber
\\ &\ &
\times\left(\frac{M_X^2}{M_{XX}}\right)
\sqrt{\frac{1}{M_{XX}^2-4M_X^2}}\ \nonumber
\\ &\ & \times\ln
\left[\frac{M_X}{\Gamma_X} \right]. \label{source1}
\end{eqnarray}

Where $M_X$ and $\Gamma_X$ are the mass and width of either $Z$,
$W$ or $t$, $M_{XX}$ is the effective mass of the $ZZ$, $WW$ or
$t\overline{t}$ pair and $v$ is the Higgs vacuum expectation value
(= 246 GeV). Here, all masses and widths are expressed in
GeV/$c^2$. The minus sign indicates that the mass of particle $X$
determined in the pair environment is always lower than the
corresponding mass determined for particle $X$ produced alone.
Note that, in the notation of \cite{Theory}, particles 1 and 2 are
identical and therefore $M_1 = M_2$.

This expression provides the shift in mass of particle $X$ as a
function of the effective mass of the pair of particles, $M_{XX}$.
As the ${XX}$ threshold is approached, $M_{XX}$ approaches $2
M_X$. The following three figures show the size of the effect in
the cases of $Z$, $W$ and $t$, respectively. As can be seen, the
predicted mass shifts are large and significant and they should be
detectable.

The following figure (Fig. \ref{fig:zmass}) shows the $Z$ mass
shift as a function of the effective mass of the $ZZ$ pair. The
$ZZ$ threshold is at 182.375 GeV/$c^2$ and in the region within
100 MeV/$c^2$ of this threshold the mass shift is always larger
than 1.5 GeV/$c^2$.  The effective mass of the $Z$ pair has to be
more than 17 GeV/$c^2$ above threshold before the $Z$ mass shift
goes below 100 MeV/$c^2$.

\vspace{0.1in}
\begin{figure}[htbp]
\begin{center}
\hspace*{0.5mm}\mbox{\epsfig{file=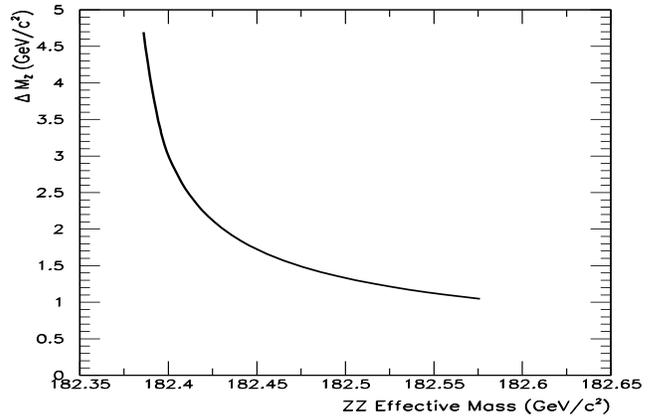,width=0.5\textwidth,height=0.24\textheight,clip=}}
\end{center}
\caption{The $Z$ mass shift plotted against the $ZZ$ effective
mass in the $ZZ$ threshold region.}\label{fig:zmass}
\end{figure}
\vspace{0.1in}

\begin{figure}[htbp]
\begin{center}
\hspace*{0.5mm}\mbox{\epsfig{file=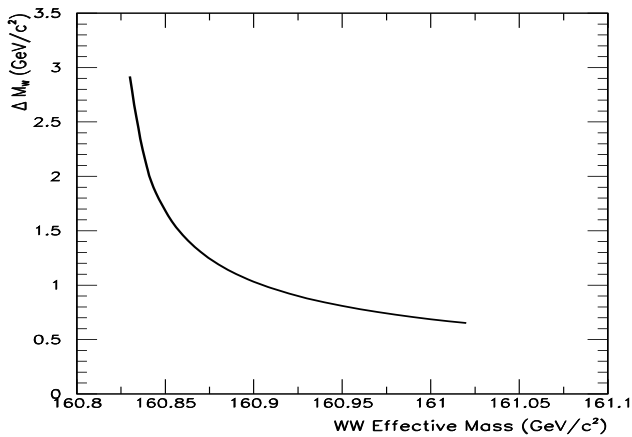,width=0.5\textwidth,height=0.25\textheight,clip=}}
\end{center}
\caption{The $W$ mass shift plotted against the $WW$ effective
mass in the $WW$ threshold region.}\label{fig:wmass}
\end{figure}

The $W$ mass shift as a function of the effective mass of the $WW$
pair is presented in Figure \ref{fig:wmass}. The $WW$ threshold is
at 160.820 GeV/$c^2$ and in the region within 100 MeV/$c^2$ of
this threshold the mass shift is always larger than 0.9 GeV/$c^2$.
The value of $M_{WW}$ has to be almost 8 GeV/$c^2$ above threshold
before the $W$ mass shift goes below 100 MeV/$c^2$.

\vspace{0.1in}
\vspace{0.1in}

\begin{figure}[htbp]
\begin{center}
\hspace*{0.5mm}\mbox{\epsfig{file=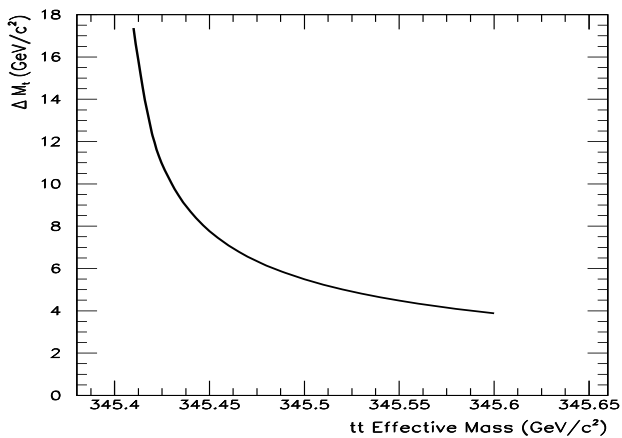,width=0.5\textwidth,height=0.25\textheight,clip=}}
\end{center}
\caption{The $t$ mass shift plotted against the $t\overline{t}$
effective mass in the $t\overline{t}$ threshold
region.}\label{fig:tmass}
\end{figure}

The $t$ mass shift as a function of the effective mass of the
$t\overline{t}$ pair is shown in Figure \ref{fig:tmass}. The
$t\overline{t}$ threshold is at 345.4 GeV/$c^2$ and in the region
within 100 MeV/$c^2$ of this threshold the mass shift is always
larger than 5 GeV/$c^2$. In this case, $M_{t\overline{t}}$ has to
be more than 10 GeV/$c^2$ above threshold before the $t$ mass
shift goes below 500 MeV/$c^2$.

\vspace{0.1in}
\vspace{0.1in}

The mass of the $Z$-boson has been determined very precisely at
LEP1 \cite{ALEPHZ,DELPHIZ,L3Z,OPALZ}.  The results are summarized
in Table \ref{zmass}. These are all measurements made at the
$Z$-pole and the average value is $M_Z$ = 91.1875 $\pm$ 0.0021
GeV/$c^2$ \cite{LEPEWZ}.

\vspace{0.1in}
\begin{table}[htpb]\centering
\begin{tabular}{|c|c|c|} \hline
Experiment&$M_Z \  GeV/c^2$&Reference \\ \hline

ALEPH - LEP1 & 91.1893 $\pm$ 0.0031&\cite{ALEPHZ,LEPEWZ}\\

DELPHI - LEP1& 91.1863 $\pm$ 0.0028&\cite{DELPHIZ,LEPEWZ}\\

L3 - LEP1    & 91.1894 $\pm$ 0.0030&\cite{L3Z,LEPEWZ}\\

OPAL - LEP1  & 91.1853 $\pm$ 0.0029&\cite{OPALZ,LEPEWZ}\\ \hline

Average& 91.1875 $\pm$ 0.0021&\cite{LEPEWZ}\\ \hline
\end{tabular}
\caption{$Z^0$ mass determinations at the
$Z^0$-pole.}\label{zmass}
\end{table}
\vspace{0.2in}

All four LEP experiments saw clear $Z$ signals in the $ZZ$ channel
at LEP2, but none published a separate $Z$ mass determination
\cite{ALEPHZZ,DELPHIZZ,L3ZZ,OPALZZ}. However, all four experiments
are consistent with the LEP2 $Z$ mass being the same as the LEP1
$Z$ mass.

\begin{figure}[htbp]
\begin{center}
\hspace*{0.5mm}\mbox{\epsfig{file=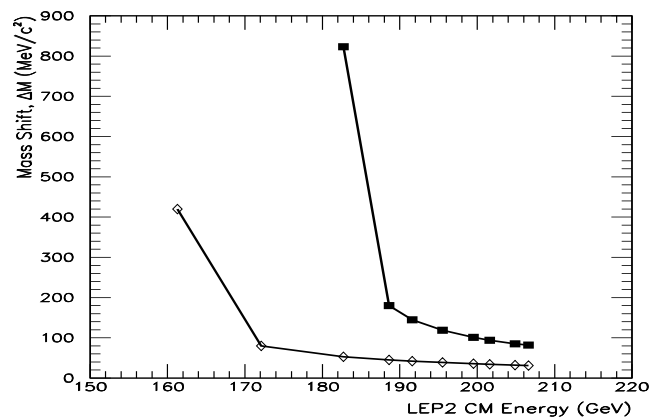,width=0.5\textwidth,height=0.25\textheight,clip=}}
\end{center}
\caption{The $Z$ (squares) and $W$ (diamonds) mass shifts plotted
against the LEP center-of-mass energy.}\label{fig:lep2masses}
\end{figure}

Figure \ref{fig:lep2masses} shows the expected $Z$ mass shift
(upper curve) as a function of the LEP2 center-of-mass energy
($E_{CM}$).  As noted earlier, $\Delta M_Z$ is larger than 100
MeV/$c^2$ until $E_{CM}$ reaches $\sim$200 GeV; the weighted
average for the $Z$ mass shift over the entire LEP2 energy range
is $\sim$140 MeV/$c^2$. It would be very interesting to see a
$Z$-candidate mass plot for the lowest energy data point at
$E_{CM}$ = 182.7 GeV where the mass shift is predicted to be
greater than 800 MeV/$c^2$ and $M_Z \sim 90.3$ GeV/$c^2$.

\vspace{0.1in}
\vspace{0.1in}

The mass of the $W$-boson has been determined quite precisely at
LEP2 \cite{ALEPHWW,DELPHIWW,L3WW,OPALWW}. All of these
measurements have been made in an environment where the $W$ is one
of a $WW$ pair. Table \ref{wmass} summarises the current
situation. The average of these measurements gives $M_W$ = 80.392
$\pm$ 0.039 GeV/$c^2$ \cite{LEPEWW}. The mass of the $W$-boson has
also been determined quite precisely at the Tevatron collider
\cite{CDFWW,DZEROWW}, where it is produced alone, and Table
\ref{wmass} again summarises the situation. Here the average of
the $W$ mass measurements is $M_W$ = 80.452 $\pm$ 0.059 GeV/$c^2$
\cite{LEPEWWG}. These two $M_W$ determinations give the same
result within errors and they can be combined to give the global
average, $M_W$ = 80.410 $\pm$ 0.032 GeV/$c^2$, presented in Table
\ref{wmass} \cite{LEPEWW}. In light of Figure
\ref{fig:lep2masses}, it is perhaps not surprising that the LEP2
and Tevatron determinations of $M_W$ are in good agreement (within
the $\sim$70 MeV/$c^2$ combined error); the expected $W$ mass
shift is below 50 MeV/$c^2$ for most of the LEP2 data and the
first two energy points, where the predicted mass shift is large,
constitute $<2\%$ of the total data set. Again it would be very
interesting to see the value of $M_W$ determined for the lowest
energy data point at $E_{CM}$ = 161.3 GeV where the mass shift is
predicted to be greater than 400 MeV/$c^2$ and $M_W \sim 80.0$
GeV/$c^2$.

%The Particle Data Group \cite{PDG} has a slightly different
%approach from that of the LEP Electroweak Working Group
%\cite{LEPEWWG} but arrive at the same conclusion. They use the
%same statistical procedure for combining the data, but consider
%only the published data of the six collaborations referenced in
%Table \ref{wmass}
%\cite{ALEPHWW,DELPHIWW,L3WW,OPALWW,CDFWW,DZEROWW}. They quote
%$M_W$ = 80.425 $\pm$ 0.038 GeV/$c^2$.

\vspace{0.1in}
\begin{table}[htpb]\centering
\begin{tabular}{|c|c|c|} \hline
Experiment&$M_W \  GeV/c^2$&Reference \\ \hline

ALEPH - LEP2 & 80.379 $\pm$ 0.058&\cite{ALEPHWW,LEPEWW}\\

DELPHI - LEP2& 80.404 $\pm$ 0.074&\cite{DELPHIWW,LEPEWW}\\

L3 - LEP2    & 80.376 $\pm$ 0.077&\cite{L3WW,LEPEWW}\\

OPAL - LEP2  & 80.416 $\pm$ 0.053&\cite{OPALWW,LEPEWW}\\ \hline

LEP2 Average & 80.392 $\pm$ 0.039&\cite{LEPEWW}\\ \hline

%UA2 - SP$\overline{\rm P}$S&80.36 $\pm$ 0.37&\cite{UA2}\\

CDF - Tevatron Collider    &80.433 $\pm$ 0.079&\cite{CDFWW}\\

D\O\ - Tevatron Collider   &80.483 $\pm$ 0.084&\cite{DZEROWW}\\
\hline

P$\overline{\rm P}$ Collider Average&80.452 $\pm$
0.059&\cite{LEPEWWG}\\ \hline

Global Average& 80.410 $\pm$ 0.032&\cite{LEPEWW}\\ \hline
\end{tabular}
\caption{$W^\pm$ mass determinations in the $W^\pm$ pair
environment.}\label{wmass}
\end{table}
\vspace{0.2in}

The mass of the $W$ has been determined in other situations where
it is not part of a pair. These are all indirect determinations.
In neutrino-nucleon scattering experiments, the most significant
of these is from a careful measurement of sin$^2\theta_W$ by the
NuTeV collaboration \cite{NUTEVW} and, assuming the value of $M_Z$
from Table \ref{zmass}, it gives $M_W$ = 80.136 $\pm$ 0.084
GeV/$c^2$. The LEP Electroweak Working Group has also determined
$M_W$ from a global standard model fit to the SLD data, LEP1 data
and the best measurement of $M_t$ \cite{LEPEWWG}. They quote $M_W$
= 80.364 $\pm$ 0.021 GeV/$c^2$.

The $t$ mass has been determined by CDF \cite{cdft} and D\O\
\cite{dzerot} and the combined average value \cite{combinedtop} is
$M_t = 172.7 \pm 2.9 \ {\mathrm{GeV}}/c^2$. The $t$'s are
presumably produced in pairs. There is no determination to date of
$M_t$ in an environment where the $t$ is produced alone. The
current status of top quark measurements from the Tevatron
Electroweak Working Group can be found at \cite{tevewwg}. It would
be interesting to investigate very carefully the events where the
$t\overline{t}$ effective mass is close to threshold: for example,
at 10, 20, 30, 40 and 50 GeV/$c^2$ above threshold, the $t$ mass
is predicted to be shifted down by 530, 360, 290, 240 and 200
MeV/$c^2$, respectively.

The calculations of reference \cite{Theory} indicate that the
Standard Model, with no extensions, predicts that masses of
particles can depend on whether or not they are produced singly or
in pairs. This paper demonstrates that the predicted mass
differences can be large near threshold and suggests that for
events in which the kinematics is well-enough defined, there may
be interesting constraints on Higgs physics from $ZZ$, $WW$ and
$t\overline{t}$ production. In particular, the $ZZ$ channel could
provide a powerful test of the prediction, especially with one $Z$
decaying to $e^+e^-$ and the other to $\mu^+\mu^-$ with all
4-vectors well measured. Even for a few fortuitous events close to
threshold one might hope to see a significant mass shift. With the
full data set, the mass shift trend might be visible in a plot of
$M_Z$ as a function of $M_{ZZ}$.

For $WW$ events and $t\overline{t}$ events there are additional
complications due to electromagnetic and strong interactions
\cite{Theory,earlierpaper}, but again the data should be looked at
with a mind open to the possibility of mass shifts.

Because the leading part of the effect is {\em independent of
Higgs mass}, failure to see the predicted effect could rule out
the Higgs mechanism in the Standard Model for all possible Higgs
masses. Observation of the mass shift would provide strong
evidence for the idea of dynamically generated masses via coupling
to the Higgs field.

\section{Acknowledgements}

We would like to thank our colleagues on LEP, Tevatron and LHC
experiments, and the NSF and INFN for their continued and generous
support.

\end{document}